\documentclass[]{aastex631}

\shorttitle{Pleiades: unresolved binaries}
\shortauthors{Malofeeva et al.}

\graphicspath{{./}{figures/}}

\begin{document}

\title{Unresolved Binaries in the Intermediate Mass Range in the Pleiades Star Cluster
\footnote{Released on November, 6th, 2021}}

\correspondingauthor{Giovanni Carraro}
\email{giovanni.carraro@unipd.it}

\author{Alina A. Malofeeva}
\affiliation{Ural Federal University \\
19 Mira Street, 620002 Ekaterinburg, Russia}

\author[0000-0001-8669-803X]{Anton F. Seleznev}
\affiliation{Ural Federal University \\
19 Mira Street, 620002 Ekaterinburg, Russia}

\author[0000-0002-0155-9434]{Giovanni Carraro}
\affiliation{Dipartimento di Fisica e Astronomia, Universita'  di Padova \\
Vicolo Osservatorio 3, I35122, Padova, Italy}

\begin{abstract}

The identification of binary stars of different mass ratios in resolved stellar populations is a challenging task.
We show how the photometric diagram constructed with the pseudo-colors (H-W2)-W1 vs W2-(BP-K) can be employed to estimate the binary and multiple star ratios and the distribution of their component mass ratio $q$ effectively.
As an application, we investigate the Pleiades star cluster in the range of primary component mass between 0.5 and 1.8 $M_{\odot}$.
The binary star ratio is found to be between 0.54$\pm$0.11 and 0.70$\pm$0.14.
On the other hand, the ratio of systems with multiplicity more than 2 is between 0.10$\pm$0.00 and 0.14$\pm$0.01.
The distribution of the component mass ratio $q$ has been approximated by a power law with the exponent between -0.53$\pm$0.10 and -0.63$\pm$0.22.
Below 0.5 $M_{\odot}$, we expect a large number of brown dwarfs among secondary components.

\end{abstract}


\section{Introduction} \label{sec:intro}

Unresolved binary and multiple systems play a crucial and yet largely underrated role in a variety of astrophysical contexts.
As an example, one needs to take them into account when constructing the initial mass function \citep{Kroupa&Jerabkova2018} of stellar systems.
Besides, the parameters of binary and multiple stars population set important constraints to the star formation theory \citep{Torniamenti+2021,Raju+2021,Cournoyer-Cloutier+2021}.
Several detailed theoretical studies have shown that the distributions of the binary and multiple star parameters do not vary significantly during the dynamical evolution of open star clusters \citep{Geller+2013,Parker&Reggiani2013}.
This implies that observations of the present-day binaries even in the oldest open clusters can  bring essential information on the primordial binary population.
Binary and multiple stars are important in many aspects of star cluster dynamical study as well \citep{Kaczmarek+2011,Geller&Leigh2015,Rastello+2020}.
The evolution of stars in close binary systems leads to the formation of `exotic' stars: blue straggler stars, cataclysmic variable stars and so forth \citep{Hurley+2001,Carraro&Seleznev2011,Hong+2017}.
Finally, one needs to take unresolved binary and multiple stars into full account when obtaining photometric estimates of open cluster masses \citep{Borodina+2019,Rastello+2020,Borodina+2021} or spectroscopic estimates of their dynamical mass \citep{Seleznev+2017,Rastello+2020}.

\begin{figure}
    \centering
    \includegraphics[width=10cm]{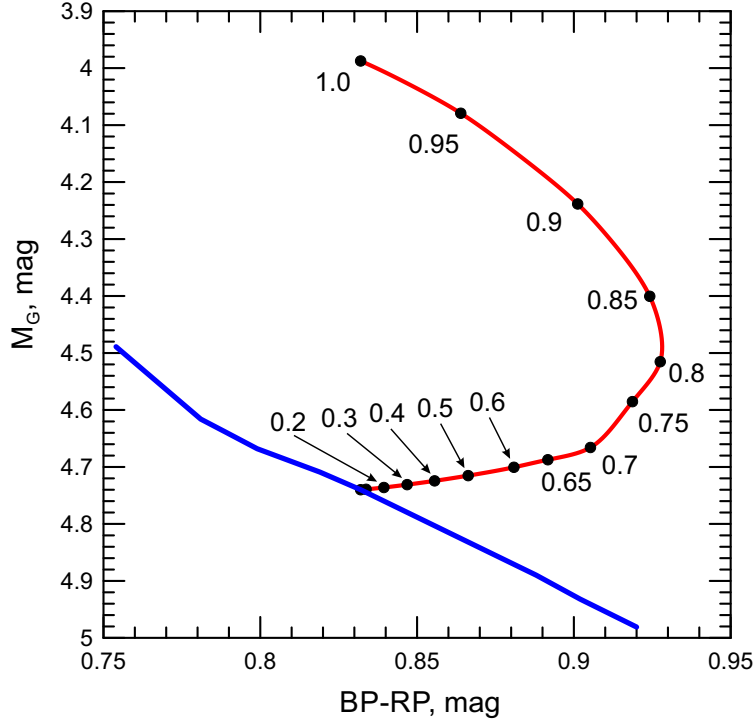}
    \caption{Deviation of the unresolved binary system from the single star sequence as a function of the $q$ value in a typical CMD in the Gaia EDR3 \citep{GaiaEDR3} photometric system.  $ M_1=1~ M_{\odot}$ .} 
     \label{binary_deviation}
\end{figure}

The search and characterization of unresolved binaries in star clusters is a tantalizing task.
If conducted via spectroscopic campaigns it typically requires the use of medium and large size telescopes and a significant investment of observing time to monitor the radial velocity time evolution.
On the other side, photometry is more appropriate.
However, photometric data in the visible wavelengths have strong limitations, since they can help to select confidently unresolved binary systems only for binary component ratio $ q=M_2/M_1\gtrsim0.4-0.5$. This is illustrated in Fig.\ref{binary_deviation}.

The currently available data on the properties of the populations of binary and multiple stars in clusters are incomplete and often contradictory.
In globular clusters the binary ratio has been found to be  $\alpha\leqslant0.1$ \citep{Milone+2012}.
However, \citet{Li+2017} had found $ \alpha=0.6-0.8$ for NGC6362, NGC6652, and NGC6838.
In open clusters the binary ratio is $\alpha\geqslant0.3$ (see \citet{Borodina+2019} for a review, and \citet{Niu+2020}).
The binary ratio for the field stars seems to be even larger (see \citet{Duchene&Kraus2013} for a review).\\

On the other hand, the literature data on the higher multiplicity systems for open clusters is  scant and, unfortunately, very discordant.
\noindent
Before reviewing it, we introduce some formalism for the binary fraction parameter $\alpha$ that, following \citet{Borodina+2021}, can be defined as:

\begin{equation}
\label{alpha}
\alpha=\frac{N_{binaries}+N_{triples}+N_{quadruples}+...}{N_{singles}+N_{binaries}+N_{triples}+N_{quadruples}+...}\; .
\end{equation}

\noindent
The dots denote systems with the multiplicity higher than four.

For the ratio of the higher multiplicity systems in the field, \citet{Tokovinin2014}  found the proportion 54:33:8:4:1.
The first number in this proportion is for single stars, the second for binary, the third for triple, the fourth for quadruple, and the fifth for quintuple systems.
We do not take into account the quintuple systems because their presence in star clusters (even open clusters) is improbable.
\citet{Borodina+2021} introduced the triples fraction $\beta$ and the quadruples fraction $\gamma$:

\begin{equation}
\label{beta}
\beta=\frac{N_{triples}}{N_{binaries}+N_{triples}+N_{quadruples}}\; ,
\end{equation}

\noindent
and

\begin{equation}
\label{gamma}
\gamma=\frac{N_{quadruples}}{N_{binaries}+N_{triples}+N_{quadruples}}\; .
\end{equation}

The result of \citet{Tokovinin2014} corresponds to $\alpha=0.45$, $\beta=0.18$, and $\gamma=0.09$, since we do not consider quintuple systems.
With these definitions, available data can be summarised as follows.\\
\citet{Mermilliod+1992} obtain for F5-K0 stars in the central part of Pleiades (a circle with radius of 70 arcminutes) the proportion 56:30:2 for singles through triples which corresponds to $\alpha=0.36$, and $\beta=0.06$.
Here and below, we take into account in formulas (2) and (3) only the existing systems.
\citet{Bouvier+1997} give for the K-dwarfs in the central part of Pleiades the proportion 119:22:3 for singles through triples which corresponds to $\alpha=0.17$, and $\beta=0.12$.
\citet{Danilov2021} has found for Pleiades in a wide mass range the proportion (260-270):(89-98):9 for {\bf singles and} both unresolved and visual binaries and triples.
This would  correspond to $\alpha=0.27-0.28$, and $\beta=0.08-0.09$.
For Praesepe, finally,  \citet{Mermilliod&Mayor1999} had found 47:30:3 for singles through triples that was $\alpha=0.41$, and $\beta=0.09$.

Diverse results are available on the mass ratio $q$ distribution as well.
\citet{Li+2020} and \citet{Danilov2021} have found  a flat $q$ distribution over the mass range of roughly 0.47-1.7 $M_{\odot}$ and for $q>0.2$ \citep{Li+2020} and for $>0.47$ solar masses \citep{Danilov2021}.

\citet{Fisher+2005,Maxted+2008,Raghavan+2010,Danilov2021} revealed a local maximum near $q=1$.
\citet{Kouwenhoven+2009} reported a power law $\frac{dN}{dq}\sim q^p$, and \citet{Reggiani&Meyer2013} found $p=0.25\pm0.29$.
Finally, \citet{Kouwenhoven+2009} provided a different expression for the $q$ distribution, namely the Gaussian one $\frac{dN}{dq}\sim \exp{-\frac{(q-\mu_q)^2}{2\sigma_q^2}}$ with $\mu_q=0.23$ and $\sigma_q^2=0.42$.
All these studies are based on data for stars of different mass range.
\citet{Fisher+2005} investigated spectroscopic binaries in the  solar neighbourhood with $M_V\leqslant4$.
Results from \citet{Raghavan+2010} and \citet{Kouwenhoven+2009} are for solar-type stars; \citet{Reggiani&Meyer2013} have compared solar-type stars and M dwarfs and have not found a significant difference.
\citet{Maxted+2008} investigated low-mass stars and brown dwarfs in young star clusters. 

\begin{figure}
    \centering
    \includegraphics[width=10cm]{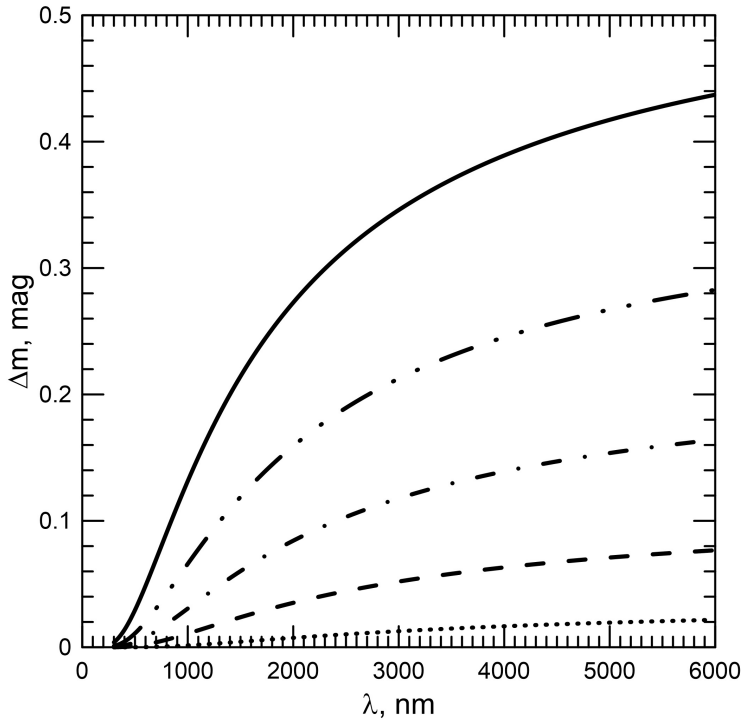}
    \caption{The differences of monochromatic stellar magnitudes of a binary and its primary component for binary stars adopting the Planck approximation. $M_1=1 M_{\odot}$. Dotted line corresponds to $q=0.1$, dashed line --- $q=0.2$, dashed-with-one-dot line --- $q=0.3$, dashed-with-two-dots line --- $q=0.4$, solid line --- $q=0.5$.  }
     \label{deltamag} 
\end{figure}

Finally, recently \citet{Thompson+2021} introduced a new method and used the spectral energy distributions (SEDs) for stars combining photometric data in several bands from visual to near infrared wavelengths.
With this method, they could unravel unresolved binary systems and determine the masses of the components.\\

\noindent
In this study, we decided to follow a different path, namely to identify that photometric diagram where single and unresolved binary stars would be most effectively separated.
To anticipate the results, this is achieved by exploring several combinations of pseudo colors, from the ultraviolet to the mid infrared wavelengths.
Our goal is to increase the precision and sensitivity to low $q$ values of unresolved binaries and to apply and test the method with the Pleiades cluster in order to be able to extend it to other clusters in the future.  

Therefore, the layout of the paper is as follows.
Section 2 is devoted to the search of promising photometric diagrams.
Section 3 describes an application of our method to the Pleiades cluster.
Finally, section 4 is dedicated to a summary of our results.

\section{A photometric diagram to separate single star from binary stars} \label{sec:diagram}

The proposed approach has a solid theoretical foundation because the binary star has a small infrared excess compared to its primary component.
Fig.\ref{deltamag} illustrates the differences of monochromatic stellar magnitudes of binary and its primary component $$m_1-m_{bin}=-2.5\log{\frac{4\pi R_1^2 J_1}{(4\pi R_1^2 J_1+4\pi R_2^2 J_2)}} \; ,$$ \noindent where $J$ is the Planck spectral density of radiation for several values of $q$ and the primary mass of $M_1=1 M_{\odot}$.
We use here the Planck approximation for the stars' spectral energy distributions to illustrate roughly the difference between a single star and a binary for different wavelengths.
Further we use more accurate results of the stellar atmosphere modeling from the isochrone tables of \citet{Bressan+2012}.
The values of an effective temperatures are from \citet{Bressan+2012}, and the values of stellar radii are obtained with the formula from \citet{Schweitzer+2019}.
It is important to underline that in the near/mid infrared range ($\leq 10\mu m$) the binary stars cannot be confused with possible debris disks which show the peak of their infrared excess at much longer wavelengths \citep{Bryden+2006}.

This implies that one needs to exploit a combinations of magnitudes in the visible pass-bands (where the SED of the binary star is very close to the SED of the  single star) and infrared pass-bands (where binary star distinguish better from the single star) in order to maximize the change to get binaries well separated from single stars. A photometric diagram with such combination of pseudo colors would then need to be searched for.

With the goal of identifying such diagram we make use of an open cluster model consisting of 200 single stars and 200 binary stars.
We use the Salpeter mass function for both single stars and the primary components of the binary stars  (in the mass range of 0.1-6.8 $M_{\odot}$) and a flat distribution for the component mass ratio $q$.
The masses of the secondary components are then selected in the mass range from 0.1 $M_{\odot}$ to $M_1$.
Unfortunately, we could not extend the mass range to  smaller masses for secondary components due to the limited mass range of the Padova suite of isochrones \citep{Bressan+2012}\footnote{http://stev.oapd.inaf.it/cgi-bin/cmd}, that we opted to use.

Then we compiled a catalog of stellar magnitudes for each star of the model in the pass-bands U, B, V, R, u, g, r, i, z, G, BP, RP, J, H, K$_S$, W1, and W2 using the above cited set of isochrones \citep{Bressan+2012}.
As a first step, we used an isochrone of $5\cdot10^7$ years.
The magnitudes of the unresolved binaries are then calculated using the following formulas where the suffix 1 refers to the primary component and the suffix 2 refers to the secondary component of the binary; moreover, $F$ refers to the radiation flux:

\begin{equation}
\label{binary}
\begin{array}{l}
m_1-m_2=-2.5\log{\frac{F_1}{F_2}}\; ,\\
x=\frac{F_1}{F_2}=10^{-0.4(m_1-m_2)}\; ,\\
m_{bin}-m_1=-2.5\log{\frac{F_1+F_2}{F_1}}\; ,\\
m_{bin}=m_1-2.5\log{(1+\frac{1}{x})}\; .
\end{array}
\end{equation}

\begin{figure}
    \centering
    \includegraphics[width=18cm]{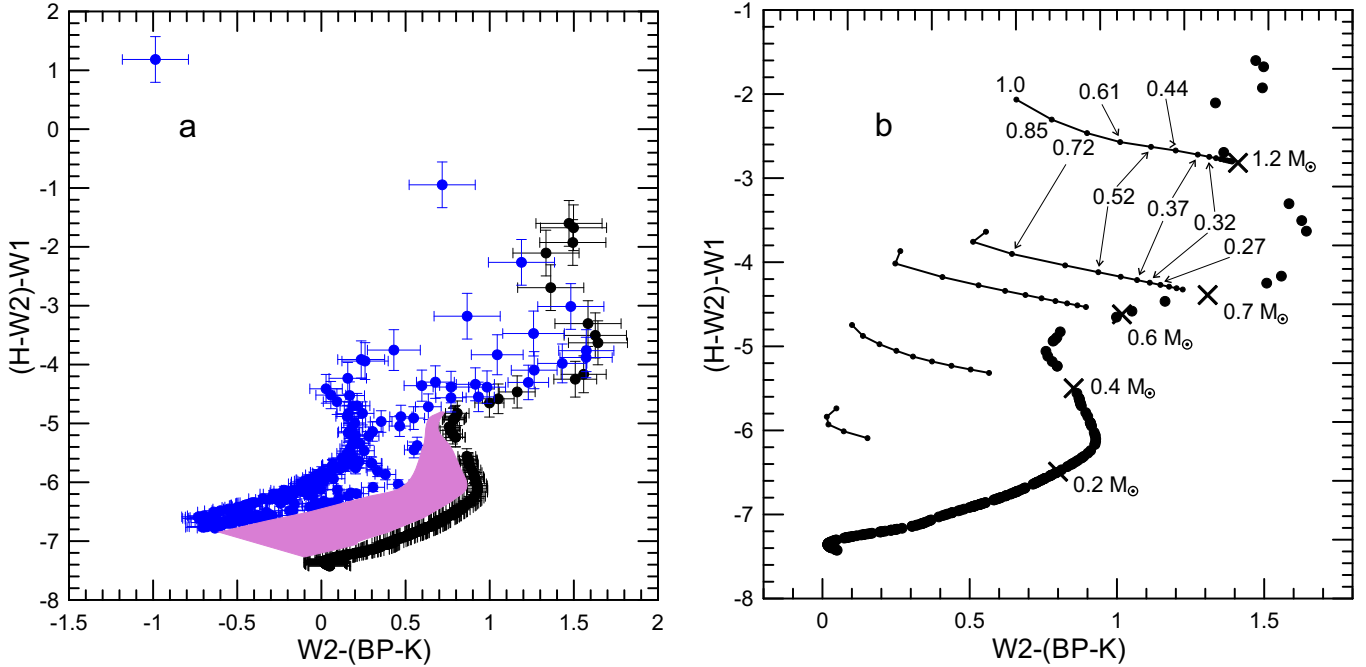}
    \caption{The diagram W2-(BP-K) vs (H-W2)-W1 for the model cluster. (a) Blue points show the binary stars, black points show the single stars. The magenta region shows an area where the model does not have binary stars due to the lack of information on stellar magnitudes of objects with masses lower than 0.1 solar masses. (b) Lines of the deviation of the unresolved binary star from the single star sequence for different masses of the primary component. Different values of $q$ are indicated by numbers and arrows.   }
     \label{model_diagram}
\end{figure}

Then, we searched for the best diagram by visually  inspecting a large number of filter combinations using the stellar magnitudes from our model catalog.
As a result, we eventually selected the photometric diagram which employs the pseudo-colors  W2-(BP-K) vs (H-W2)-W1. This diagram is shown in Fig.\ref{model_diagram}.
Error bars in Fig.\ref{model_diagram}a show the typical errors of the pseudo-colors calculated with the use of the random errors of magnitudes in the pass-bands BP, H, K, W1, W2 from \citet{Lodieu+2019} for Pleiades cluster.
In order to apply these random errors, we use the mean dependencies `magnitude - photometric error' for these pass-bands plotted by the data from \citet{Lodieu+2019} catalog.

We emphasize that the lines of deviation of the binary star sequence from the single star sequence are nearly horizontal.

The gap between the single stars and binary stars sequences in the lower part of the diagram in Fig.\ref{model_diagram}a (indicated in magenta) corresponds to binary stars with the secondary component having a mass lower than 0.1 solar mass.
This gap is due to the lack of data on such objects in the tables of theoretical isochrones of \citet{Bressan+2012}.

In order to exploit this diagram to estimate the binary ratio in  real clusters, we need to use an isochrone table corresponding to the cluster age. In order to evaluate $q$ ratios, we plot lines of the constant $q$ values.

\begin{figure}
    \centering
    \includegraphics[width=18cm]{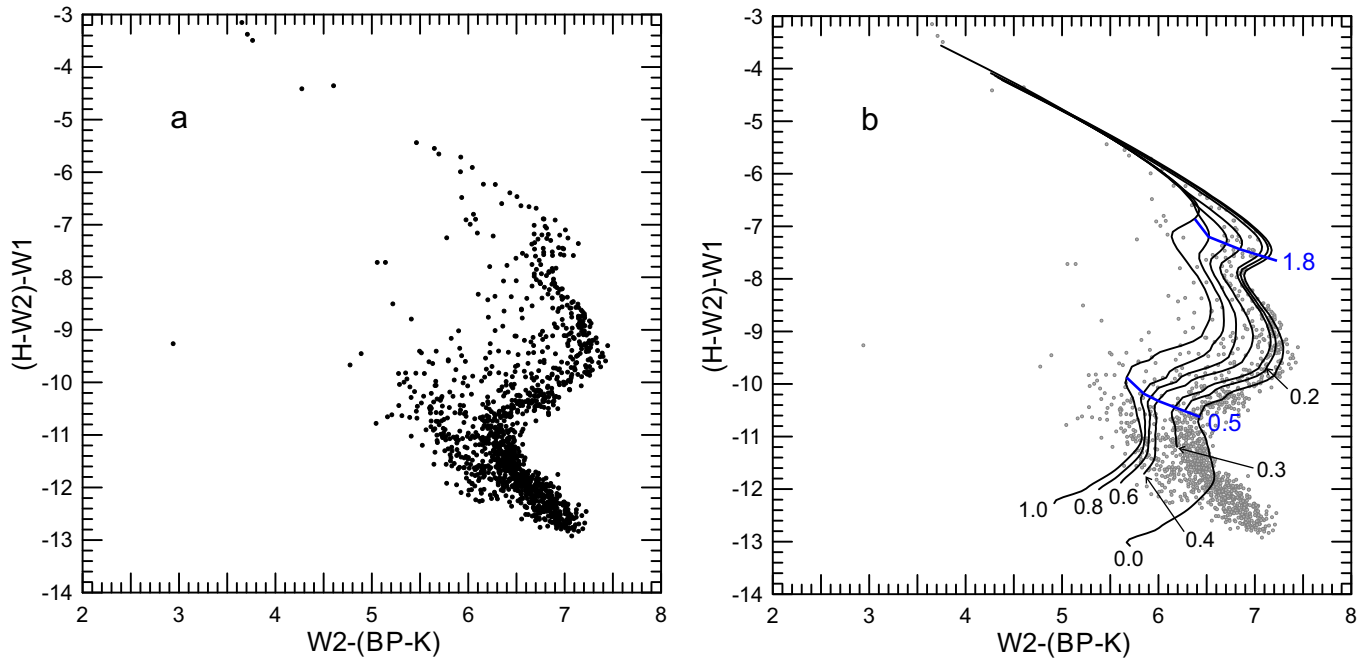}
    \caption{(a) The diagram W2-(BP-K) vs (H-W2)-W1 for the combined sample of Pleiades probable members. (b) Lines of the constant $q$ values are over-imposed onto the Pleiades stars' distribution. The black numbers are $q$ values, the blue lines and numbers correspond to masses of primary components. The lines of the constant $q$ values correspond to the \citet{Bressan+2012} isochrone with $\bf \log{t}=8.1$.}   
     \label{Pleiades_diagram}
\end{figure}

\section{The binary population of the Pleiades cluster} \label{sec:Pleiades}

In order to study the binary and multiple star population of the Pleiades star cluster, we compiled a sample of the most probable cluster members. 
This sample is the combination of two data-sets. The first one consists of  1391 stars obtained by \citet{Danilov&Seleznev2020} inside a circle corresponding to the cluster radius of 10.9 degree for the magnitude $G<18$ mag. It includes stars with  membership probability larger than 95\% and it is complete up to the $\sim$90\% level \citep{Danilov&Seleznev2020}.
The second one is from \citet{Lodieu+2019}.
The catalog of \citet{Lodieu+2019} contains all stellar magnitudes necessary to build up the aforementioned photometric diagram.

The diagram W2-(BP-K) vs (H-W2)-W1 for the combined sample is shown in panel a) of Fig.\ref{Pleiades_diagram}.
In panel b) of the same figure lines of the constant $q$ values are overlapped on the Pleiades star distribution.
To this aim, we employed the Pleiades fundamental parameters from the catalog of \citet{Dias+2021}, namely, the distance 135 pc, logarithm of age 8.1, $A_V=0.168$, and a solar metallicity (\citet{Dias+2021} give $ [Fe/H]=0.032\pm0.029$) .\\

\noindent
A few interesting conclusions can be drawn from this figure.

\begin{itemize}
\item this diagram is clearly not suitable in the massive stars range since the constant $q$ value lines overlap;
\item the isochrone from \citet{Bressan+2012} does not follow the cluster sequence in the lower part of the cluster diagram for the primary component mass lower than 0.5 solar mass; one reason could be these stars are pre-main-sequence stars; another reason could be the isochrones of \citet{Bressan+2012} are not suited for low-mass stars due to some intrinsic limitation;
\item  we do not see any gap on the left side of the single star sequence. This might mean that the Pleiades harbour a lot of unresolved binaries with very-low-mass secondary components, most probably brown dwarfs.
\end{itemize}

\begin{figure}
    \centering
    \includegraphics[width=10cm]{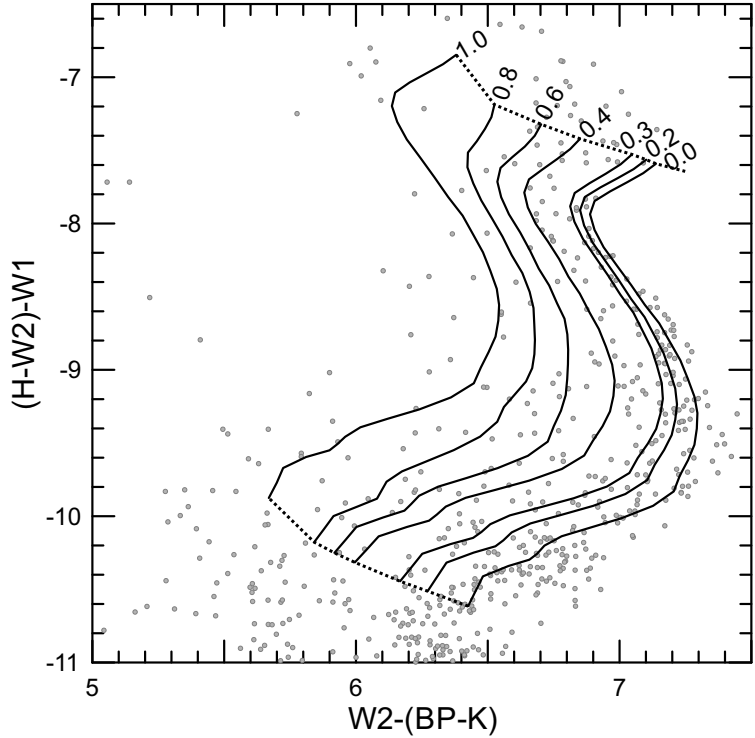}
    \caption{The diagram W2-(BP-K)--(H-W2)-W1 for the counts of stars with different values of $q$. Lines of the constant $q$ values are shown. }
     \label{Pleiades_counts}
\end{figure}

\noindent
For the reasons outlined above, we can use our diagram only in the intermediate range of the primary component masses, namely, from 0.5 to 1.8 $M_{\odot}$.
We estimated the number of unresolved binaries with different values of $q$ by star counts within the diagrams, an illustration of which is provided in Fig.\ref{Pleiades_counts}.
In order to take into account photometric uncertainties, we  perform star counts five times as described below.

The first time we used the stellar magnitudes from the input catalog  as they are given there (Fig..\ref{Pleiades_counts}).
Next, we displaced all stars by the value of individual errors of photometric indices four times: up and down (by the value of an error of (H-W2)-W1) to the left and to the right (by the value of an error of W2-(BP-K)).
To do this, the one-sigma errors of the photometric indices (H-W2)-W1) and W2-(BP-K) were calculated according to the one-sigma errors of magnitudes that make up each index.
We calculated star numbers between lines of the constant $q$ values, to the left of the line $q=1$ and to the right of the line $q=0$.
The results of star counts are shown in Table 1.

\begin{deluxetable*}{cc}
\tablenum{1}
\tablecaption{The results of counts of stars with different values of $q$\label{tab:counts}}
\tablewidth{0pt}
\tablehead{
\colhead{The range of $q$} & \colhead{Number of stars} \\
}
\startdata
to the right of $q=0$ & 122$\pm$59  \\
$0<q<0.2$             & 67$\pm$12   \\
$0.2<q<0.3$           & 42$\pm$11   \\
$0.3<q<0.4$           & 73$\pm$15   \\
$0.4<q<0.6$           & 29$\pm$5    \\
$0.6<q<0.8$           & 21$\pm$6    \\
$0.8<q<1.0$           & 22$\pm$1    \\
to the left of $q=1.0$& 29$\pm$4    \\
\enddata
\end{deluxetable*}

We consider the stars to the left side of the line $q=1.0$  as unresolved multiple stars with multiplicity larger than 2.
It is more difficult to evaluate the number of single stars.
We attempted  this in two ways.
Firstly, we consider stars to the right side of the line $q=0$ as single {\bf stars}.
Secondly, we added stars from the interval $0<q<0.2$ to the number of single stars.
In the first case we obtain $\alpha=0.70\pm0.14$ and $\beta+\gamma=0.10\pm0.00$. In the second case we obtain $\alpha=0.54\pm0.11$ and $\beta+\gamma=0.14\pm0.01$.

Fig.\ref{Pleiades_counts} starts to show a systematic variation for stars to the right of the q=0 line at about -8.4 on the vertical axis.
There are two reasons for this.
Firstly, it is a joint effect of increasing the number of stars and of the errors of photometric indices whose typical values are shown in Fig.\ref{model_diagram}a.
Secondly, it is due to partial match of the isochrone with the cluster sequence.
This variation can  increase the false single star counts artificially.
However, this is partially accounted for by our count procedure because the typical values of the indices errors are comparable with the star count variations.

\begin{figure}
    \centering
    \includegraphics[width=10cm]{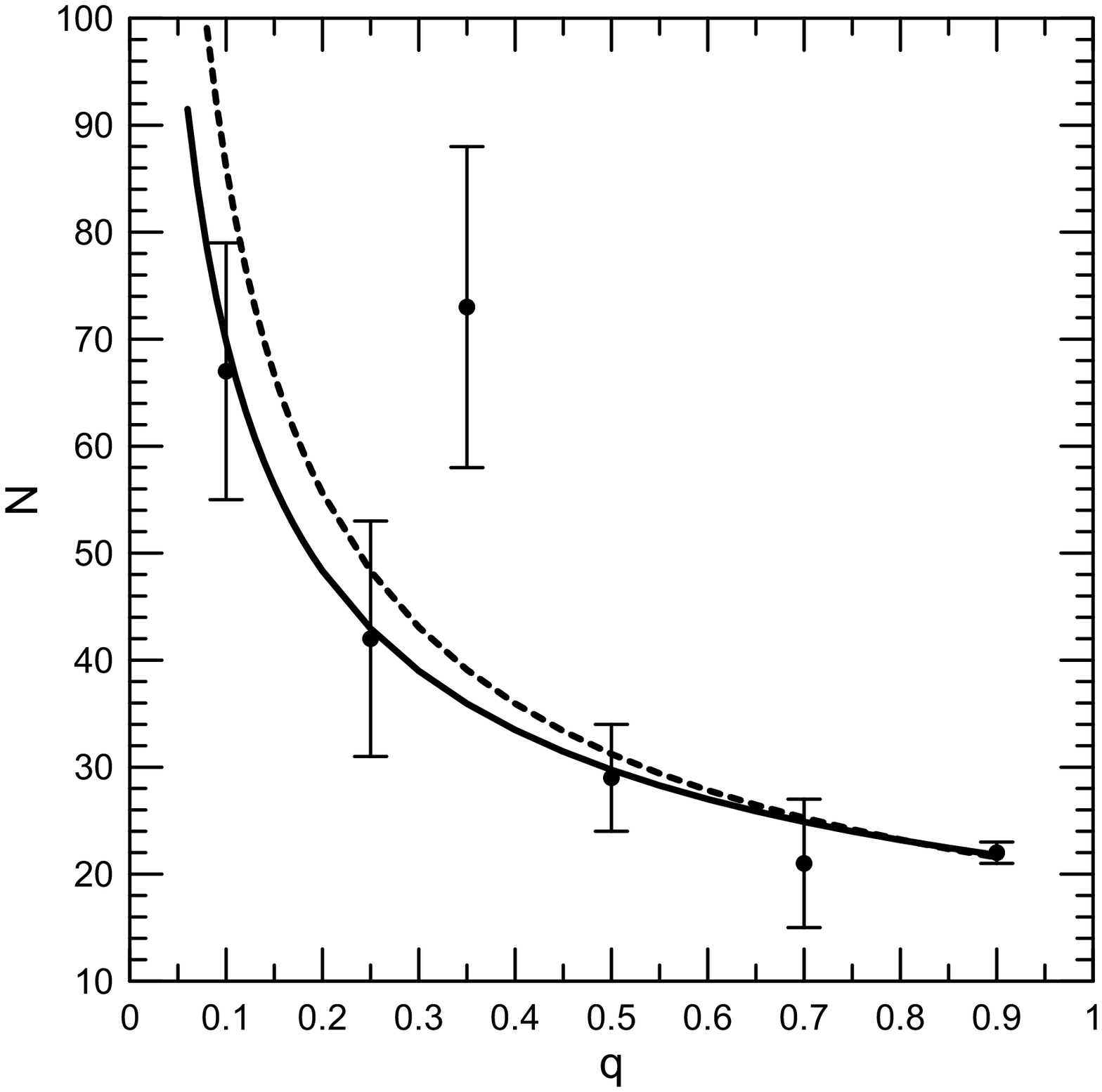}
    \caption{The distribution of the component mass ratio $q$ and its approximation by a power law. Solid line --- stars with $0<q<0.2$ are considered as binaries; dashed line --- stars with $0<q<0.2$ are considered as singles.  }
     \label{q_distr}
\end{figure}

The distribution of $q$ is shown in Fig.\ref{q_distr}.
We approximated this distribution with a power law as a result of a non-linear least squares method using the curve\_fit procedure from the Python package scipy.optimize.
The point for $q=0.35$ deviates from this dependence indicating that the number of unresolved binaries with small $q$ could be even larger.
In the first case we obtain $f(q)=(20.60\pm1.70)\cdot q^{-0.53\pm0.10}$ (solid line in Fig.\ref{q_distr}).
In the second case we obtain $f(q)=(20.18\pm2.06)\cdot q^{-0.63\pm0.22}$ (dashed line in Fig.\ref{q_distr}).

\section{Conclusions} \label{sec:Conclusions}

In this study, we investigated the population of  binary and multiple stars in the Pleiades star cluster.
To this purpose, we identified a photometric diagram which employ two pseudo-colors constructed from stellar magnitudes in the pass-bands of visible and infrared wavelengths W2-(BP-K) vs (H-W2)-W1.
Binary stars in this diagram pop clearly up and are well distinguished from the single stars even for small values of the component mass ratio $q$.

Unfortunately, we limited the exploration of this diagram to intermediate masses for the primary component ($m_1\in[0.5;1.8]$ $M_{\odot}$), because of technical constraints.

In the future we plan to expand this investigation both to more massive stars (another diagram should be found) and to less massive stars (we need to explore different isochrone sets to reproduce better the main sequence of the Pleiades star cluster in the low-mass regime).

Unfortunately, it is problematic to derive parameters of triple and, possibly, quadruple systems basing on our diagram, because the position of unresolved star beyond the region of binaries on the diagram could be reproduced in numerous ways. However, additional spectroscopic and photometric observations of these systems can bring more clarity.

The comparison of  the diagram W2-(BP-K) vs (H-W2)-W1 for Pleiades probable members and for the cluster model indicates that in the low-mass range the Pleiades star cluster harbours numerous binary stars with very-low-mass secondary components.
We expect that in general these secondary components are brown dwarfs.

In the considered mass range the binary ratio is between $\alpha=0.54\pm0.11$ and $\alpha=0.70\pm0.14$.
The ratio of the multiple stars with multiplicity greater than 2 is between $\beta+\gamma=0.10\pm0.00$ and $\beta+\gamma=0.14\pm0.01$.
These estimates are larger than previously found (see above in introduction).
We can easily account for  this result since our diagram allows one to detect unresolved binaries with smaller component mass ratio which were clearly missed in previous investigations.

The large number of unresolved binaries with small values of $q$ is contained in the $q$ distribution itself. We can approximate it by a power law with the exponent between $-0.63\pm0.22$ and $-0.53\pm0.10$.

As a by-product of our investigation we make available a list of the probable Pleiades members which are candidate unresolved binary stars with the estimate of their component mass ratio as well.
This list can be exploited for future detailed spectroscopic  and photometric studies of these stars.

Further, we plan to apply this method to other open clusters to see if the parameters of unresolved binaries population are the same from cluster to cluster, which could help to elucidate their formation and  dynamical evolution. 

\begin{acknowledgments}
The work of A.F.Seleznev was supported by the Ministry of Science and Higher Education of the Russian Federation, FEUZ-2020-0030.The work of G. Carraro has been supported by Padova University grant BIRD191235/19: {\it Internal dynamics of Galactic star clusters in the Gaia era: binaries, blue stragglers, and their effect in estimating dynamical masses}.
\end{acknowledgments}

\bibliography{Pleiades_ubms_1}{}
\bibliographystyle{aasjournal}



\end{document}